\documentclass{pnastwo}

\usepackage{PNAStwoF}
\usepackage{amssymb,amsfonts,amsmath}
\usepackage{graphicx}
\newcommand{\mb}[1]{\mathbf{\displaystyle #1}}
\newcommand{\bs}[1]{\boldsymbol{\displaystyle #1}}

\contributor{Submitted to Proceedings of the National Academy of Sciences of the United States of America}
\url{www.pnas.org/cgi/doi/10.1073/pnas.0709640104}
\copyrightyear{2008}
\issuedate{Issue Date}
\volume{Volume}
\issuenumber{Issue Number}
\begin{document}

%% For titles, only capitalize the first letter
\title{Dynamic network drivers of seizure generation, propagation and termination in human epilepsy}

\author{Ankit N. Khambhati\affil{1}{Department of Bioengineering, University of Pennsylvania, Philadelphia, PA 19104, USA}
    \affil{2}{Penn Center for Neuroengineering and Therapeutics, University of Pennsylvania, Philadelphia, PA 19104, USA},
    Brian Litt\affil{1}{}
    \affil{2}{}
    \affil{3}{Department of Neurology, Hospital of the University of Pennsylvania, Philadelphia, PA 19104, USA},
    \and
    Danielle S. Bassett
    \affil{1}{}
    \affil{2}{}
    \affil{4}{Department of Electrical and Systems Engineering, University of Pennsylvania, Philadelphia, PA 19104, USA}}

\contributor{Submitted to Proceedings of the National Academy of Sciences of the United States of America}

\maketitle

%%%%%%%%%%%%%%%%%%%%%%%%%%%%%%%%%%%%%%%%%%%%%%%%%%%%%%%%%%%%%%
\begin{article}

\begin{abstract}
Drug-resistant epilepsy is traditionally characterized by pathologic cortical tissue comprised of seizure-initiating `foci'. These `foci' are thought to be embedded within an epileptic network whose functional architecture dynamically reorganizes during seizures through synchronous and asynchronous neurophysiologic processes. Critical to understanding these dynamics is identifying the synchronous connections that link foci to surrounding tissue and investigating how these connections facilitate seizure generation and termination. We use intracranial recordings from neocortical epilepsy patients undergoing pre-surgical evaluation to analyze functional connectivity before and during seizures. We develop and apply a novel technique to track network reconfiguration in time and to parse these reconfiguration dynamics into distinct seizure states, each characterized by unique patterns of network connections that differ in their strength and topography. Our approach suggests that seizures are generated when the synchronous relationships that isolate seizure `foci' from the surrounding epileptic network are broken down. As seizures progress, foci reappear as isolated subnetworks, marking a shift in network state that may aid seizure termination. Collectively, our observations have important theoretical implications for understanding the spatial involvement of distributed cortical structures in the dynamics of seizure generation, propagation and termination, and have practical significance in determining which circuits to modulate with implantable devices.
\end{abstract}


\abbreviations{ECoG, electrocorticography}

\subsection{Significance Statement}
Localization-related epilepsy affects $\approx$80\% of epilepsy patients and is often resistant to medication. The challenge for treating patients is mapping dynamic connectivity between cortical structures in the epileptic network during seizures. While it is well known that whole-brain functional architecture reconfigures during tasks, we hypothesize that epileptic networks reconfigure at the meso-scale leading to seizure initiation, propagation, and termination. We develop new methods to track dynamic network reconfiguration amongst connections of different strength as seizures evolve. Our results indicate that seizure onset is primarily driven by the breakdown of strong connections that re-surge in an isolated focal sub-network as seizures transition to termination. These findings have practical implications for targeting specific connections with implantable, therapeutic devices to control seizures.
\linebreak

\dropcap{L}ocalization-related epilepsy is traditionally characterized by seizures that arise from one or more abnormal islands of cortical tissue in the neocortex or mesial temporal structures, such as hippocampus \cite{siegel2001medically}. In more severe cases, seizures with focal onset secondarily generalize, as pathologic activity spreads across the brain \cite{kutsy1999ictal}. Localization-related epilepsy represents $\approx$80\% of epilepsy cases and is often resistant to medication \cite{french2007refractory}. For drug-resistant patients, the only treatment options are implantable devices, or more traditionally resective surgery to remove enough cortical tissue in the epileptic network to decrease seizure frequency, while preserving brain tissue responsible for eloquent function. In surgical cases where discrete lesions associated with seizure onset (`foci') are not evident on brain MRI, only $\approx$40\% remain seizure-free post-surgery \cite{french2007refractory}. The modest outcome associated with these procedures has lead investigators to further explore spatial distributions of epileptic activity using multiscale neural signals in ECoG and sub-millimeter $\mu$ECoG to more accurately localize where seizures start and how their pathologic activity spreads \cite{worrell2008high-frequency, schevon2009spatial, stead2010microseizures, viventi2011flexible, feldt_muldoon2013spatially, weiss2013ictal}. These approaches have spurred a paradigm shift from localizing epileptic `foci' towards mapping the epileptic network and identifying key regions for intervention.

In a broad sense ``functional networks'' describe time-dependent communication pathways between neural populations. They can be measured by metrics that establish statistical relationships between sensor dynamics. In this framework sensors (nodes) that exhibit a significant degree of synchrony are considered functionally connected. The notion of an epileptic network stems from the idea that pathologic synchrony manifests as disruptions in neural function: rhythmic motor activity, altered cognition, or abnormal sensation. While seizures were originally thought to be limited to bursts of hypersynchrony (strong connections between nodes), recent work demonstrates that seizures are characterized by more complex temporal dynamics demonstrating synchrony and asynchrony in tandem \cite{jerger2005multivariate, schindler2006assessing, schindler2008evolving, kramer2010coalescence, jiruska2012synchronization}.

Clinical epileptologists and translational researchers are anxious to employ these methods to answer pressing clinical questions: Where do seizures start? Can the epileptic network be modulated therapeutically? What can these methods reveal about the underlying neurophysiologic mechanisms? Answering these questions requires the ability to track time-dependent functional connections \cite{holme2012temporal} and to study the geometry of epileptic networks as seizure-generating regions interact with the surrounding cortex. Such a capability is vital, because epileptogenic regions cause symptoms not only through their own dysfunction, but also through their ability to recruit and disrupt normal brain regions.

Based on the complexity of seizure dynamics, we hypothesize that network reconfiguration explains how seizures begin, spatially progress and self-terminate. In this view, epileptic network reconfiguration forms a causal prediction of activity, rather than a post-hoc description. Our hypothesis is informed by recent work demonstrating that human brain networks dynamically reconfigure prior to changes in behavior \cite{bassett2011dynamic,ekman2012predicting}. These reconfigurations can be characterized by variations in strong (synchronous) and weak (asynchronous) connections, which have been proposed in prior work to support different network functions \cite{ercsey-ravasz2013predictive, santarnecchi2014efficiency}. We hypothesize that the epileptic network can be characterized by strong connections representing primary propagation pathways, and weak connections that initiate seizure onset; these two populations uniquely define the topographical extent and propagation dynamics of seizures.

\section{Results}
\begin{center}
\begin{figure}[t]
\includegraphics{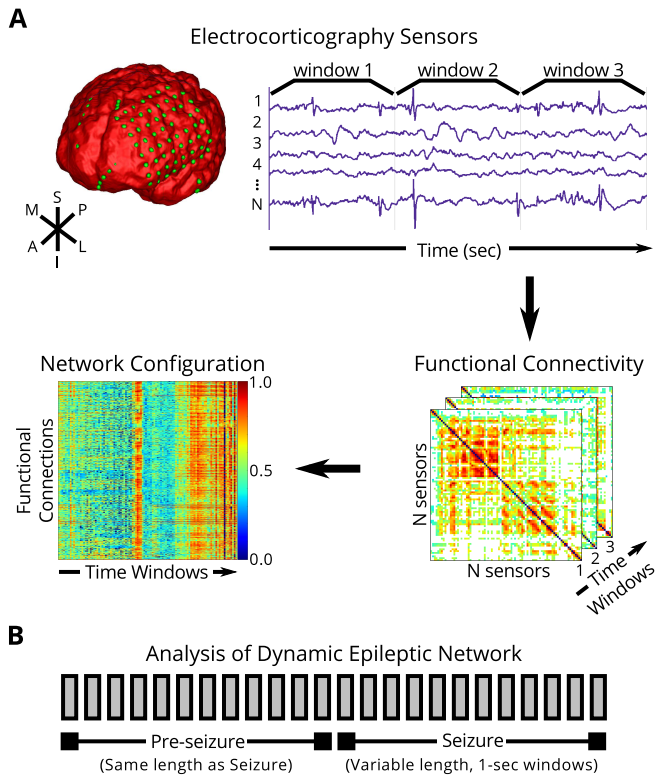}
\caption{\textbf{Schematic of Analysis Pipeline for Dynamic Epileptic Networks.} (\textit{A}) (\textit{Top}) We create functional networks based on electrophysiology by windowing ECoG signals collected from patients with medically refractory epilepsy implanted with intracranial electrodes into 1s time windows. Each sensor is represented as a network node, and weighted functional connectivity between sensors, interpreted as degree of synchrony, is represented as a network connection. (\textit{Lower Right}) Functional connectivity is estimated by a magnitude cross-correlation between sensor time series for each time window; non-significant connections are identified via comparison to surrogate data \cite{bassett2013robust} and removed (see \textit{Materials and Methods}). (\textit{Lower Left}) We visualize the resultant data as a connection-by-time matrix that represents all connection weights for all 1s time windows. (\textit{B}) For each seizure, we form a dynamic seizure network from the set of time windows that range from the earliest electrographic change characterizing seizure onset \cite{litt2001epileptic} to the seizure termination point. Similarly, we form a dynamic pre-seizure network from a set of time windows extracted directly prior to seizure onset. Time window size in both epileptic and pre-seizure networks are identical.\label{fig1}}
\end{figure}
\end{center}

To address these hypotheses, we examined ECoG sensor-level activity prior to and during seizure epochs. We assessed sensor-sensor functional connectivity by constructing dynamic network representations from independent time windows of data (Fig.~\ref{fig1}A) for both seizure and pre-seizure periods (Fig.~\ref{fig1}B). In each 1s time window of ECoG data, the sensors were represented as network nodes and inter-sensor magnitude cross-correlations, or synchrony, were represented as weighted network connections.

\subsection{Functional Network Architecture Reveals Distinct Seizure States}
\begin{center}
\begin{figure}[b]
\includegraphics{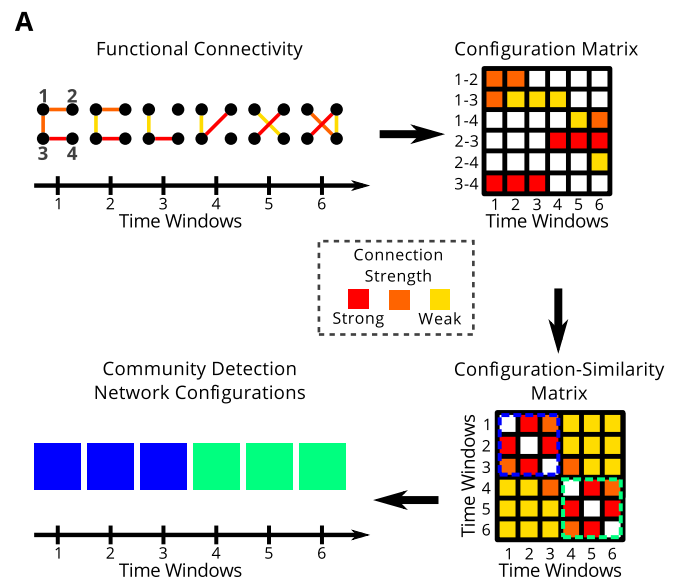}
\caption{\textbf{Identifying Network Configuration States.} (\textit{First}) We construct dynamic functional networks; colors represent arbitrary connection strengths. (\textit{Second}) We track functional connectivity over time using a connection-by-time matrix, or `configuration matrix', that represents the set of connection weights for each 1s time window. (\textit{Third}) We compute a Pearson correlation between all possible pairs of time windows to construct a configuration-similarity matrix; colors represent the magnitude of similarity and visually identified clusters are distinguished by colored, dashed lines. (\textit{Fourth}) We apply community detection to the configuration-similarity matrix to extract clusters of time windows with similar network configurations; colors represent assignments of time windows to different network configuration communities.\label{fig2}}
\end{figure}
\end{center}

Do functional connectivity patterns change as a seizure progresses? To answer this question, we developed a new method to uncover network states, defined by unique patterns of sensor-sensor functional connectivity between $T$ time windows (Fig.~\ref{fig2}). We summarized the time-dependent fluctuations in connectivity between all sensors in a configuration matrix (see \textit{Materials and Methods}). To quantify similarity in gross network connectivity between time windows, we calculated the Pearson correlation between pairs of time windows from the configuration matrix, thereby forming a symmetric $T \times T$ configuration-similarity matrix (see \textit{Materials and Methods}). We cluster the configuration-similarity matrix to group time windows with similar network connectivity patterns into distinct network states, which we discuss below.

\begin{center}
\begin{figure*}[t]
\includegraphics{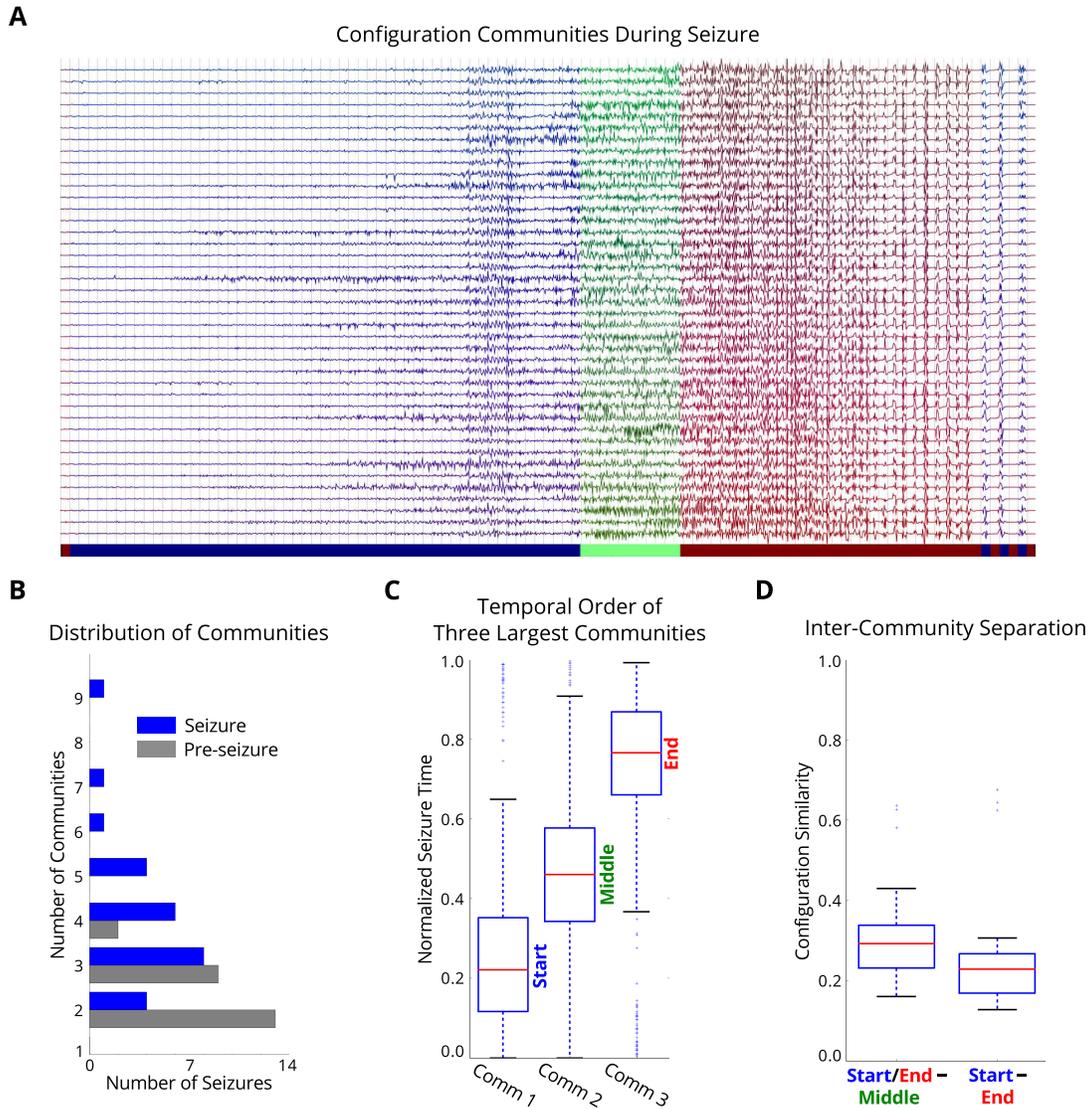}
\caption{\textbf{Distinct States of Dynamic Epileptic Networks.} (\textit{A}) Example assignments of sensors to network configuration communities (seizure states) demonstrating network reconfiguration. Assignments are overlaid on a subset of representative ECoG grids during an entire seizure. (\textit{B}) Number of communities in pre-seizure (gray) and seizure (blue) epochs; $t$-test ($N=25$, $t=0.50$, $p\approx0.62$). (\textit{C}) Distribution of community assignments to temporal time windows for seizure epochs; we normalize all seizure lengths to 1. (\textit{D}) Average separation of the \emph{start} and \emph{end} communities from the \emph{middle} community compared to separation between \emph{start} and \emph{end} over seizure population---connection strengths averaged between all nodes of each community; $t$-test ($N=21$, $t=3.40$, $p<0.005$).\label{fig3}}
\end{figure*}
\end{center}

To test for distinct network states across seizure epochs, we used an unsupervised network clustering approach known as community detection. This detection method maximizes a modularity quality function $\mb{Q}$ obtained from the configuration-similarity matrix (Fig.~\ref{fig2}; for details, see \textit{Materials and Methods}). When applied to individual seizure epochs from each patient, this procedure revealed ``configuration communities.'' Each community was composed of a group of time windows with similar network configurations, and therefore corresponded to a unique brain state (Fig.~\ref{fig3}A). The number of communities in pre-seizure networks and seizure networks were not statistically different ($N=25$, $T=0.50$, $P\approx0.62$) (Fig.~\ref{fig3}B). However, seizure communities were less randomly distributed in time, indicating that they captured meaningful temporal structure (see \textit{SI} for measuring temporal dispersion). Seizure networks tended to display 3 large communities (see \textit{SI} for distilling communities); which we refer to as \emph{start}, \emph{middle} and \emph{end} communities, based on their temporal placement in the seizure (Fig.~\ref{fig3}C).

Next we asked whether clustered network configurations within these three seizure states were distinct or whether certain pairs of states shared more or less similarity than other pairs of states. To address the question of state separability, we calculated the correlation between connectivity patterns of different time windows from the configuration matrix, in which each time window was assigned a different configuration community. We found that connectivity patterns in \emph{start} and \emph{end} communities were less similar to one another, on average (Fig.~\ref{fig3}D). In contrast, connectivity patterns in the \emph{middle} community were more similar to both \emph{start} and \emph{end} communities, suggesting that the \emph{middle} community plays an intermediary role as a transition state.

\subsection{Seizure States Display Stereotyped Connectivity}
\begin{center}
\begin{figure}[b]
\includegraphics{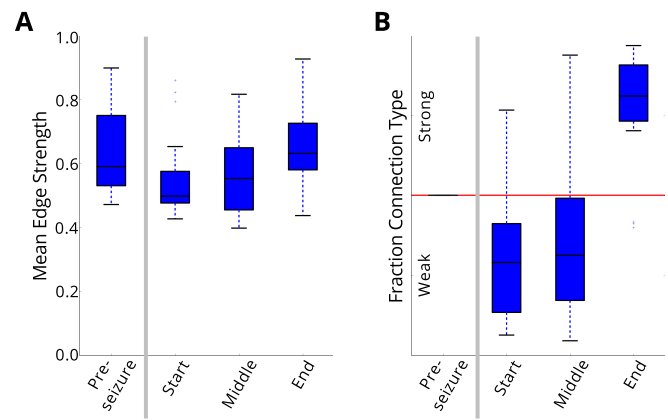}
\caption{\textbf{Network Geometry of Seizure States}. (\textit{A}) Distribution of connection strengths over seizure population---connections strengths averaged over all time windows within each community compared to pre-seizure epoch; one-way ANOVA ($N=21$, $df=3$, $F=3.41$, $p<0.05$). (\textit{B}) Distribution of the fraction of strong to weak connections over seizure population---$B$ within each community compared to pre-seizure epoch; one-way ANOVA ($N=21$, $df=3$, $F=25$, $p<1\times10^{-13}$). Strong (weak) connections were determined separately for each epoch based on the top (bottom) 10\% of the connection weight distribution. See \textit{Materials and Methods} for a formal definition of $B$.\label{fig4}}
\end{figure}
\end{center}

In the previous section, we observed that seizures progress through distinct states characterized by different patterns of sensor-sensor functional connectivity. To understand how these patterns differ, we used a two-pronged approach, examining (i) the average and (ii) the distribution of functional connectivity in each time window as a function of seizure state (Fig.~\ref{fig4}). The mean functional connectivity (averaged over all pairs of sensors and over all time windows in a community) increased as the seizure progressed through the 3 states (Fig.~\ref{fig4}A), indicating that the gross epileptic network becomes increasingly synchronized.

To investigate the distribution of connection strengths, we separately tracked \emph{strong} and \emph{weak} connections through each epoch, defined as the 10\% largest valued and 10\% smallest valued connections in each community, respectively. To measure the balance between these connections in each state, we defined a balance index $\mb{B}$ (see \textit{Materials and Methods}): values approaching $+1$ indicate that most connections are strong while values close to $-1$ indicate that most connections are weak. Pre-seizure epochs exhibited a balance of connection type (Fig.~\ref{fig4}B). In seizure epochs, the \emph{start} community was dominated by weak connections, while the \emph{end} community was dominated by strong connections. The \emph{middle} community displayed a large variance in the balance index over seizures, consistent with its putative role as a transition state between \emph{start} and \emph{end} communities. These results suggest that a finer examination of network geometry in the \emph{middle} community might help to explain how brain networks transition through seizure states.

\subsection{Anatomical Structure of Dynamic Epileptic Networks}
\begin{center}
\begin{figure}[b]
\includegraphics{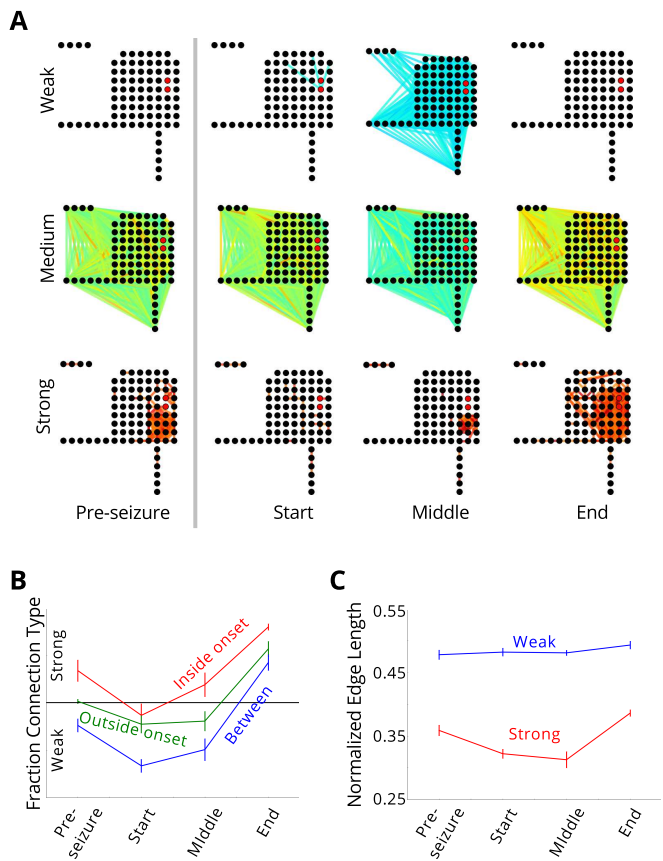}
\caption{\textbf{Dynamics of Network Geometry.} (\textit{A}) Example of network geometry with preserved 2-D spatial relationships between nodes for categories of \emph{weak} ($<10\%$), \emph{medium} ($10$-$90\%$) and \emph{strong} ($>90\%$) connections; topographic map represents mean connection strength within a seizure community or pre-seizure epoch. The clinically-determined seizure onset sensors are shown in red. (\textit{B}) Population average balance index, $B$, for three connection locations (within seizure-onset zone, outside seizure-onset zone, and between seizure-onset zone and other sensors) across epochs and communities. (\textit{C}) Physical length of connections in a large $8x8$ grid, normalized to one for longest possible connection, averaged over connections and seizures for each configuration community. Error bars indicate standard error of the mean over seizures.\label{fig5}}
\end{figure}
\end{center}

In the preceding analyses, we demonstrated that seizures progress through distinct states characterized by different network geometries (patterns of weighted connections). However, whether these reconfigurations are spatially localized or distributed, and how they relate to seizure foci, was unclear. To address these questions, we leveraged routine clinical procedures: A team of neurologists identified the sensors on the seizure onset zone based on visual inspection of the intracranial recordings. We used this information to map connections in each seizure state to physical electrode locations in stereotaxic space. Throughout pre-seizure and seizure epochs, we observed that clusters of \emph{strong} connections lay in close proximity to seizure onset sensors (Fig.~\ref{fig5}A).

To quantify this observation and examine its role in seizure propagation, we delineated the following three topographical connections within the seizure network: (i) inside the seizure onset zone, (ii) outside the seizure onset zone, and (iii) between seizure onset sensors and all other sensors. We performed a two-way analysis of variance (ANOVA) with location and community as categorical factors and the balance index, $\mb{B}$, as dependent variable (Fig.~\ref{fig5}B). Significant main effects of location ($F=32.02$, $df=2$, $p\approx5\times10^{-13}$) and community ($F=55.40$, $df=3$, $p\approx4\times10^{-27}$) indicated that anatomical substrates of the epileptic network and temporal state play distinct roles in dynamic functional connectivity patterns. We did not observe a significant interaction effect ($F=0.86$, $df=6$, $p\approx0.52$). During pre-seizure epochs, sensors in the seizure onset zone predominantly displayed strong connections, while sensors in the surrounding epileptogenic cortex displayed weak connections. Strong connections increased from \emph{start} to \emph{middle} states within the seizure onset zone, and between the seizure onset zone and surrounding epileptogenic cortex. In contrast, strong connections linking sensors in the surrounding epileptogenic cortex did not increase until the \emph{end} state. This pattern of results is consistent with a propagation of strong connectivity beginning in the seizure onset zone and emanating to the surrounding cortex towards the end of the seizure.

The change in anatomical location of synchronous activity suggests that connections change in physical length throughout the seizure. We observed that weak connections tended to be relatively long, and remained approximately unchanged in length in the pre-seizure and seizure epochs (Fig.~\ref{fig5}C). In contrast, strong connections tended to be relatively short. The length of strong connections in the \emph{start} and \emph{middle} states was shorter than in the pre-seizure state, suggesting a more anatomically focal distribution at seizure onset. As the seizure progressed from \emph{middle} to \emph{end} states, strong connections grew longer, suggesting a broadening of the anatomical distribution of functional coordination consistent with widespread synchronization.

\section{Discussion}

\subsection{Epileptic Network Reconfiguration}
Intuitively, complex reconfiguration of functional brain networks can accompany changes in cognitive state or changes in behavior. Prior fMRI studies have explored such reconfiguration in whole-brain networks constructed from data acquired during motor skill learning \cite{bassett2011dynamic} and as task states change \cite{bassett2006adaptive}, and in networks impacted by stroke \cite{wang2010dynamic, grefkes2011reorganization}. In contrast, here we explore the reconfiguration of a local area and use high resolution ECoG data to map the fine-scale temporal dynamics of reconfiguration processes. Our approach differs from the conventional one, focused on tracking changes in sensor signals, irrespective of inter-sensor relationships. The dynamic network framework we impose allows us to study time-dependent changes in connectivity between sensors through similarities in signal statistics.

In this study, we developed and exercised a novel method for distinguishing brain states based on differences in time-dependent functional network configurations. We applied our technique to a set of human ECoG recordings, and extracted network dynamics during seizure and pre-seizure epochs. We found that seizures most often exhibit three stereotyped network configurations and that these configurations correspond to the start, middle and end of a seizure. These data-derived states may provide a fundamental neurophysiological explanation for the neurologically-defined onset, propagation and termination states ubiquitous in clinical descriptions. However, the mechanisms supporting these 3 states differ significantly in the two definitions. Clinically, seizure propagation is believed to involve the spreading of pathologic spiking activity outside of a seizure focus \cite{kutsy1999ictal} followed by a \textit{spontaneous} shift towards seizure termination. Our results paint a different picture: the middle configuration state, clinically thought to be the seizure propagation stage, plays more of a transitional role that initiates and guides seizure termination by \textit{gradually} integrating the surrounding neocortex with the seizure onset zone through strong, synchronous, connections.

\subsection{Connectivity Imbalance}
Our analytical approach utilizes the strength of functional connections to characterize them as ``strong'' (synchronous) or ``weak'' (asynchronous), rather than simply stating that two sensors are functionally ``connected'' or ``not connected''. Mathematically, this focus corresponds to a study of network \emph{geometry} as opposed to network \emph{topology}. A primary advantage of the weighted network approach is the ability to separate connections into classes that differ in strength. Evidence suggests that strong and weak connections play different roles in supporting cognitive function \cite{schneidman2006weak, santarnecchi2014efficiency}. Strong connections have traditionally been thought to represent primary communication pathways between brain areas. However, recent work demonstrates that weak connections support increased network efficiency and may play a large role in distinguishing pathologic \cite{bassett2012altered} and healthy \cite{cole2012global, santarnecchi2014efficiency} network states. From a dynamical perspective, strong connections may be those that are engaged consistently throughout neurophysiological processes, whereas weak connections may be engaged transiently to enable transitions of the brain between distinct states.

Prior work has speculated that weak connections present in the middle of a seizure \cite{schindler2006assessing, kramer2010coalescence}. However, we believe a more accurate way to address this hypothesis is to consider the strength of functional connections. Using a weighted connectivity approach, we find that connections in epileptic networks are much weaker at the start and middle of seizures, compared to pre-seizure periods, suggesting that a distributed profile of weak connectivity accompanies seizure onset. These data-driven findings challenge the conventional clinical view that seizure onset is characterized predominantly by increased synchronization. In contrast, the weak connectivity that we observe at seizure onset may represent an inability of surrounding epileptogenic cortex to quarantine seizure initiation in the seizure onset zone.

In contrast to the start of the seizure, the middle of the seizure displays a large variability over subjects, with some seizures displaying relatively weak connectivity and others displaying relatively strong connectivity. We speculate that the disparity in dominant connection type during the middle of the seizure can be described by mixed expression of connectivity patterns from the start and end of the seizure. This explanation is corroborated by our finding that network configurations in the start and end of the seizure, which are weak and strong type dominant respectively, are statistically similar to configurations in the middle of the seizure. It follows that reconfiguration during the middle of the seizure allows the epileptic network to purge weak, asynchronous, connections from the start of the seizure and to mobilize strong, synchronous, connections in a transition to the end of the seizure.

\subsection{Functional Incorporation of Seizure Onset Sub-Network}
While temporal network structure provides rich information regarding seizure states, it does not directly provide information regarding the spatial processes involved in seizure dynamics. We therefore complemented the temporal network approach by incorporating information about sensor location in Euclidean space and sensor role either within or outside the seizure onset zone. Our results demonstrate that these additional spatial features provide new insights into potential neurophysiological mechanisms involved in seizure initiation, and may inform the development of clinical tools for objectively isolating the seizure onset zone directly from seizure or pre-seizure data.

Prior work has demonstrated the presence of low synchronization between seizure-generating regions and non-seizure generating regions in epochs at least 2 hours prior to seizure start \cite{warren2010synchrony}. However, the temporal dynamics and geometrical roles of these two sets of areas has remained elusive. Our results elucidate the role played by seizure-onset regions during seizures and the accompanying recruitment of the surrounding epileptic network during termination. Clear isolation of the onset zone exists in pre-seizure periods, suggesting the potential to identify foci, niduses of seizure generation, within the network from inter-ictal data. Critically, this isolation of the onset zone diminishes at seizure initiation. As the network transitions into the middle state, we observe a resurgence of strong connections within the onset zone, consistent with pre-seizure periods. These results lead us to posit that the incorporation of the onset zone underlies a shift from seizure initiation to termination that then requires recruitment of surrounding epileptogenic cortex. The similarity of connection strength and spatial location in pre-seizure and middle seizure states suggests a mechanistic role for functional connectivity in subduing seizure onset.

\subsection{Network Tightening During Seizures}
Our observation that strong connections are typically short and weak connections are typically long, is consistent with results from two lines of research: (i) functional studies in healthy individuals that utilize other imaging modalities such as fMRI \cite{santarnecchi2014efficiency} and (ii) structural connectivity studies in non-human primates that utilize tract tracing techniques \cite{ercsey-ravasz2013predictive}. The length of strong connections decreases from the pre-seizure period to the middle of the seizure, suggesting that strong connections are physically tightening, perhaps into more functionally cohesive portions of cortex, before the seizure begins terminating. We speculate that the tightening of strong connections to a localized sub-network might act as a control mechanism to quench disruptive network activity that may have built-up over many hours prior to the seizure through increasing frequency of epileptiform discharges \cite{litt2001epileptic}. Given this explanation we theorize that seizures are a neurophysiologic mechanism to relieve a build-up of pathologic activity, akin to a rubber band recoiling after having stretched too far. This process occurs in three-stages where: (i) seizure-initiation invokes network reconfiguration that, (ii) enables strong connections to tighten and quench pathologic activity as seizures progress, (iii) the tightening spearheads a second reorganization event that synchronizes surrounding weak connections in order to stop the mechanism.

\subsection{Clinical Impact and Future Work}
We have seen that the dynamical processes that propel epileptogenic networks into seizures can be complex and are poorly understood. Yet, clinicians rely on visual inspection to describe spatial and temporal properties of seizures. The lack of standardized clinical measures to mark epileptic events calls for the development of automated methods. The network analysis tools we have built, while generally applicable to any dynamic network, can parse seizure states, localize driver `foci' of seizures, and characterize how seizures terminate. This interpretation can be translated into useful clinical tools to identify dysfunctional anatomical regions that drive the epileptic network and may be particularly amenable to local interventions, such as surgery or device placement. Of interest, seizure driving `foci' were equally present in the half of our study patients who did not have focal lesions on brain imaging, compared to those patients with lesions demonstrated on MRI. We plan a more detailed study in the future to correlate mapping of these seizure-driving regions with brain resection and outcome. 

Currently, our tools employ community detection techniques to identify gross changes in the meso-scale architecture of network structure across time. The observed meso-scale reconfiguration processes may be accompanied by region-specific trends in reconfiguration between the epileptic network and surrounding healthy networks. This work could potentially be used to address cellular mechanisms by considering micro-scale reconfigurations. Recent studies suggest that epileptic networks in the neocortex may be composed of distributed micro-domains on the scale of a few cortical columns generating high frequency oscillations and micro-seizures that coalesce in a network during seizure generation and termination \cite{stead2010microseizures}. While of great interest, these studies are currently limited by the lack of appropriate implantable high resolution sensors capable of covering clinically relevant areas sufficiently to yield comprehensive high-resolution maps. Further development of dynamic community detection methods to identify and track reconfiguration within network sub-regions at both the meso and micro-scales may help delineate healthy and pathologic networks and uncover mechanisms of network recruitment.

An important clinical consideration related to this work is the impact of sampling error inherent in any intracranial implantation procedure on our results. Any technique used to map epileptic networks, subdural electrode strips and grids, more distributed ``Stereo EEG'' implantations, and combinations of these two approaches, usually yield incomplete representations of epileptic networks. It is not possible to fully record from the entirety of cortex in affected patients. In some cases this might mean that neither seizure onset zones nor all regions of seizure spread are fully delineated.  Despite this incomplete representation, the presence of three clear states defining seizures in each of the patients presented above, and their objective and independently determined relationship to the seizure onset zone suggest that our findings are important and real. With further validation on a larger number of patients with both lesional and non-lesional epilepsies, we hope to demonstrate the utility of our method to define functional components of epileptic networks. We believe the method can be useful both in epilepsy surgery and for placing devices into regions that drive seizure initiation and termination. We also plan to use these methods to compare competing approaches for localizing epileptic networks, such as subdural and stereo EEG.  We hypothesize that each will have advantages in recording components of epileptic networks in different types of localization-related epilepsy.

\subsection{Conclusion}
We have found evidence supporting our hypothesis that epileptic networks reconfigure through 3 different synchronous states during seizures: initiation, propagation, and termination. Gross, uniform network connectivity preceding seizures was found to redistribute primarily as weak connections during seizure initiation and propagation and strong connections during termination. This indicates that an imbalance of connectivity between strong and weak connection types plays a mechanistic role in seizure topography. These imbalances lead the epileptic network to assimilate the seizure-onset zone into the surrounding tissue during the beginning of seizures and reassert its isolation as the seizure progresses. Finally, we found that these networks physically tighten strong connections into a focal sub-network during seizures, suggesting a novel mechanism for seizure termination.

\begin{materials}
    Seven patients undergoing surgical treatment for medically refractory epilepsy believed to be of neocortical origin underwent implantation of subdural electrodes to localize the seizure onset zone after noninvasive monitoring was indeterminate. De-identified patient data was retrieved from the online International Epilepsy Electrophysiology Portal (IEEG Portal); and all patients gave written informed consent in accord with the University of Pennsylvania Institutional Review Board for inclusion in this study \cite{wagenaar2013multimodal}. ECoG signals were recorded and digitized at 500 Hz sampling rate using Nicolet C64 amplifiers and pre-processed to eliminate line noise. Surface electrode (Ad Tech Medical Instruments, Racine, WI) configurations, determined by a multidisciplinary team of neurologists and neurosurgeons, consisted of linear and two-dimensional arrays (2.3 mm diameter with 10 mm inter-contact spacing) and sampled the neocortex for epileptic foci. Signals were continuously recorded for the duration of a patient's stay in the epilepsy monitoring unit. See Table S1 for demographic and clinical information.

    We analyzed a total of twenty-five complex partial seizures, a subset of which exhibited secondary generalization. We excluded simple, partial seizures and seizures stemming from non-neocortical foci from this study. Seizure onset time and localization were defined by the point of earliest electrographic change (EEC) and annotated and marked as a part of routine clinical workup \cite{litt2001epileptic}. ECoG signal directly preceding each seizure and equal in duration to that seizure was also extracted for balanced comparison and labeled as pre-seizure.

    Signals from each epoch were divided into $1$-second, non-overlapping, wide-sense stationary time-windows in accord with other studies \cite{kramer2010coalescence}. To test the biasing effect of high-amplitude spiking on signal connectivity measurements we also investigated windows $0.5$-seconds in duration to sample more of the non-biasing temporal space and found similar results.

    Dynamic functional networks were formed by applying a normalized cross-correlation similarity function $\bs{\rho}$ between the time series of two sensors in the same time window using the formula
    \begin{eqnarray}
        \bs{\rho}_{\mb{xy}}(\mb{k}) = \underset{\tau}{\operatorname{argmax}}\:{\mathrm{E}}[(\mb{x_k}(t) - \mu_{\mb{x_k}})(\mb{y_k}(t+\tau) - \mu_{\mb{y_k}})]
    \end{eqnarray}
    where $\mb{x}$ and $\mb{y}$ are signals from one of $\mb{N}$ sensors or network nodes, $\mb{k}$ is one of $\mb{T}$ non-overlapping, one-second time windows, and $\mb{x_{k}}=\mb{y_{k}}=0$. The $\mb{N}$x$\mb{N}$x$\mb{T}$ similarity matrix $\bs{\rho}_{\mb{xy}}$ was statistically thresholded in each time window $\mb{k}$ using a two-step approach \cite{bassett2013robust}. A surrogate distribution of similarity matrices was constructed for Fourier phase-randomized versions of time series in each time window and used to compute two-tailed $p$-values for the significance of the cross-correlation between every sensor pair. Finally, a false discovery rate ($\bs{\alpha}=0.05$) technique was applied to each distribution of $p$-values to retain significant connection weights. The collection of significant, weighted dynamic connections were stored in an $\mb{N}$x$\mb{N}$x$\mb{T}$ network adjacency matrix $\mb{A}$ (Fig.~\ref{fig1}A).

    Temporal changes in connectivity between nodes was tracked within individual epochs by determining modules of the configuration-similarity matrix. We construct the configuration-similarity matrix by first unraveling $\mb{A}$ to a network evolution matrix $\mb{\hat A}$ describing the weights of $\mfrac{\mb{N}(\mb{N}-1)}{2}$ connections across $\mb{T}$ time windows. Using a Pearson correlation similarity metric, we transform $\mb{\hat A}$ to a fully-connected $\mb{T}$x$\mb{T}$ configuration state adjacency matrix $\mb{S}$. The configuration adjacency matrix is partitioned into communities by maximizing the modularity index $\mb{Q}$ \cite{newman2004finding} using a Louvain-like locally greedy algorithm \cite{blondel2008fast}. We employed a Newman-Girvan null model \cite{newman2006modularity, porter2009communities} and adaptively determined an optimal structural resolution parameter $\gamma$ per seizure (see \textit{SI}; and \cite{bassett2013robust} for a more detailed discussion of resolution parameters in modularity maximization). We used a consensus partition method with 1000 optimizations per run until we obtained consistent community partitioning \cite{lancichinetti2012consensus, bassett2013robust}. The three largest communities from each seizure were selected for further analysis and re-labeled in order of median temporal occurrence for population-level comparison.

    Connections were classified as \emph{strong} or \emph{weak} based on thresholds determined by the distribution of connection strengths for each epoch separately for each seizure. The \emph{strong} (\emph{weak}) connections must be >90\% (<10\%) of all connection strengths. Non-significant connections (disconnected) were excluded from the distribution and not categorized as \emph{weak}. To measure the dominance of \emph{strong} or \emph{weak} connections, we defined the balance index as
    \begin{eqnarray}
        \mb{B} = \frac{\mb{E_s}-\mb{E_w}}{\mb{E_s}+\mb{E_w}}
    \end{eqnarray}
where $\mb{E_s}$ and $\mb{E_w}$ are the average number of strong and weak connections over possible connections and number of time windows.

Connection topography metrics were computed for only within-grid electrodes (configurations of $8 \times 8$), ignoring all other non-grid electrodes such that inter-electrode spacing in all analyses was kept constant. To measure the topographical tightness of connections, we defined the normalized connection length $\mb{L}$ as the mean Euclidean distance of all connections scaled by the longest possible distance between two nodes.
\end{materials}

\begin{acknowledgments}
    AK and BL acknowledge support from the National Institutes of Health through awards \#R01-NS063039, \#1U24 NS 63930-01A1, the Citizens United for Research in Epilepsy (CURE) through Julie's Hope Award, and the Mirowski Foundation. DSB acknowledges support from the Alfred P. Sloan Foundation, the Army Research Laboratory through contract no. W911NF-10-2-0022 from the U.S. Army Research Office, the Institute for Translational Medicine and Therapeutics at Penn, and the National Science Foundation through award \#BCS-1441502.
\end{acknowledgments}

%\bibliographystyle{pnas.bst}
%\bibliography{references.bib}

\end{article}
%%%%%%%%%%%%%%%%%%%%%%%%%%%%%%%%%%%%%%%%%%%%%%%%%%%%%%%%%%%%%%%

%%%%%%%%%%%%%%%%%%%%%%%%%%%%%%%%%%%%%%%%%%%%%%%%%%%%%%%%%%%%%%%

%% Adding Figure and Table References
%% Be sure to add figures and tables after \end{article}
%% and before \end{document}

%% For figures, put the caption below the illustration.
%%
%% \begin{figure}
%% \caption{Almost Sharp Front}\label{afoto}
%% \end{figure}

%% For Tables, put caption above table
%%
%% Table caption should start with a capital letter, continue with lower case
%% and not have a period at the end
%% Using @{\vrule height ?? depth ?? width0pt} in the tabular preamble will
%% keep that much space between every line in the table.

%% \begin{table}
%% \caption{Repeat length of longer allele by age of onset class}
%% \begin{tabular}{@{\vrule height 10.5pt depth4pt  width0pt}lrcccc}
%% table text
%% \end{tabular}
%% \end{table}

%% For two column figures and tables, use the following:

%% \begin{figure*}
%% \caption{Almost Sharp Front}\label{afoto}
%% \end{figure*}

%% \begin{table*}
%% \caption{Repeat length of longer allele by age of onset class}
%% \begin{tabular}{ccc}
%% table text
%% \end{tabular}
%% \end{table*}

\end{document}

\maketitle

~\\
$^1$Department of Bioengineering, University of Pennsylvania, Philadelphia, PA 19104, USA \\
$^2$Penn Center for Neuroengineering and Therapeutics, University of Pennsylvania, Philadelphia, PA 19104, USA \\
$^3$Department of Neurology, Hospital of the University of Pennsylvania, Philadelphia, PA 19104, USA \\
$^4$Department of Electrical and Systems Engineering, University of Pennsylvania, Philadelphia, PA 19104, USA \\

%%%%%%%%%%%%%%%%%%%%%%%%%%%%%%%%%%%%%%%%%%%%%%%%%%%%%%%%%%%%%%

\section{Patient Demographics}
\begin{table}[H]
    \caption{Patient Demographics. Seizure types were either complex partial (CPS) or complex partial with secondary generalization (CPS+Gen). M, male; F, female.}
    \begin{tabular}{cccp{3cm}cc}
        \hline
        Patient (IEEG Portal) & Sex & Age (Onset/Surgery) & Etiology & Seizure Type (N) & Imaging\\ \hline
        \verb|I002_P002| & M & 3.5/20 & Left frontal onset, \newline left cortical dysplasia & CPS+Gen (1) & Lesional\\
        \verb|I002_P003| & M & 2/36 & Right temporal onset & CPS+Gen (3) & Unclear\\
        \verb|I002_P006| & F & 15/26 & Symptomatic meningitis, \newline right temporal onset & CPS (1), CPS+Gen (4) & Non-lesional\\
        \verb|I002_P015| & M & <1/54 & Left anterior temporal onset & CPS (5) & Lesional\\
        \verb|I002_P016| & F & <1/22 & Symptomatic meningitis, \newline occipital lobe & CPS (3) & Lesional\\
        \verb|I002_P017| & F & 36/41 & Cryptogenic left temporal lobe & CPS+Gen (6) & Non-lesional\\
        \verb|I002_P023| & F & 18/25 & Left temporal lobe & CPS+Gen (2) & Non-lesional\\
    \end{tabular}
\end{table}

\section{The Configuration Vector}
For dynamic networks there is a lack of approaches to quantify how the pattern of connections between nodes, or the gross network configuration, varies over time. In networks where $N$ nodes stay constant over all $T$ time windows, we can describe the state of $\frac{N(N-1)}{2}$ possible connections between all nodes in each time window $t=1~\mathrm{to}~T$ by a \textit{configuration vector} $\vec{v_{t}} \in \mathbb{R}^{\frac{N(N-1)}{2}}$; collectively forming a \textit{configuration matrix} $\textbf{V} \in \mathbb{R}^{\frac{N(N-1)}{2}\times T}$. Intuitively, changes in $\vec{v_{t}}$ between different $t$ can imply dynamical changes in network connectivity.

While dynamic networks commonly experience meso-scale changes in connectivity over time, there is currently no way to determine whether these meso-scale changes contribute to gross reconfiguration of the entire network. Using a similarity function (herein we use Pearson correlation), we quantify the degree of change in connectivity patterns in $\textbf{V}$ between all $T$ time windows stored in a symmetric \textit{configuration-similarity matrix} $S \in \mathbb{R}^{T\times T}$. In this work we are interested in clustering time windows that express similar network configuration patterns. To this end, we employ static community detection techniques to group time windows with similar gross network configuration.

\section{Modularity Optimization for Community Detection}
\subsection{Theory}
Community-detection is an often used technique to study meaningful group structure in complex networks by clustering nodes into `modules' `communities'. A community is defined as a set of nodes that are connected among one another more densely than they are to nodes in other communities. A popular way to identify community structure is to optimize a quality function known as modularity $\mb{Q}$, which has been shown to extract meaningful functional components of networks by comparing the real network to a null model \cite{newman2004finding, newman2006modularity, porter2009communities, fortunato2010community}. While these approaches have been mainly applied to both binarized and weighted static networks, more recent extensions of the model investigate dynamic communities that group nodes over time. In our work we apply static community detection to the configuration-similarity matrix, viewed as a fully connected graph where each node represents the similarity of configuration vector $\vec{v_{t}}$ between time windows $t_1$ and $t_2$ where $t_1=t_2=0$, to group gross network configuration patterns into communities.  

In general, to perform community detection one begins with a network of $N$ nodes and a given set of connections between those nodes. A static network is then represented using an $N \times N$ adjacency matrix $\mathbf{A}$. The element $A_{ij}$ of the adjacency matrix represents a connection from node $i$ to node $j$ where $i=j=0$, and its value indicates the weight of that connection. Then, $\mb{Q}$ is formally defined as:
\begin{equation}\label{modopt}
	Q = \sum_{ij} [A_{ij} - \gamma P_{ij}] \delta(g_{i},g_{j})\,,
\end{equation}
where node $i$ is assigned to community $g_{i}$, node $j$ is assigned to community $g_{j}$, the Kronecker delta $\delta(g_{i},g_{j})=1$ if $g_{i} = g_{j}$ and it equals $0$ otherwise, $\gamma$ is the \emph{structural resolution parameter}, and $P_{ij}$ is the expected weight of the edge connecting node $i$ to node $j$ under a specified null model. The choice $\gamma = 1$ is very common, but it is important to consider multiple values of $\gamma$ to examine groups at multiple scales (see \textit{SI Section \ref{subsec:structres}}). A commonly chosen null model is the Newman-Girvan null model \cite{porter2009communities, fortunato2010community, newman2004finding, newman2006modularity} defined as:
\begin{equation}\label{newgirv}
    P_{ij} = \frac{k_i k_j}{2m}
\end{equation}
where $k_i=\sum_j A_{ij}$ is the strength of node $i$ and $m=\frac{1}{2}\sum_{ij}A_{ij}$ is the average strength over all connections. 

Maximizing $Q$ partitions the network into communities such that the total within-community connection weight is maximum relative to the null.

\subsection{Structural Resolution Parameter for Community Detection}
\label{subsec:structres}
The structural resolution parameter $\gamma$ of the modularity quality function can be used to identify community structure over different topological or geometric scales \cite{reichardt2006statistical, fortunato2007resolution, porter2009communities, mucha2010community, bassett2013robust}. Typically, $\gamma$ is chosen to be equal to 1. However, by varying $\gamma$, one can examine fine grain structure in networks by extracting smaller communities. In our application, we were interested in determining community structure of configuration-similarity matrices with a variable (i) number of sensors implanted per patient, (ii) seizure duration and (iii) spatial extent of epileptogenic cortex. Each of these factors can potentially impact the scale at which appreciable meso-scale functional reorganization can be detected.

\begin{figure}[b]
    \includegraphics[width=\textwidth]{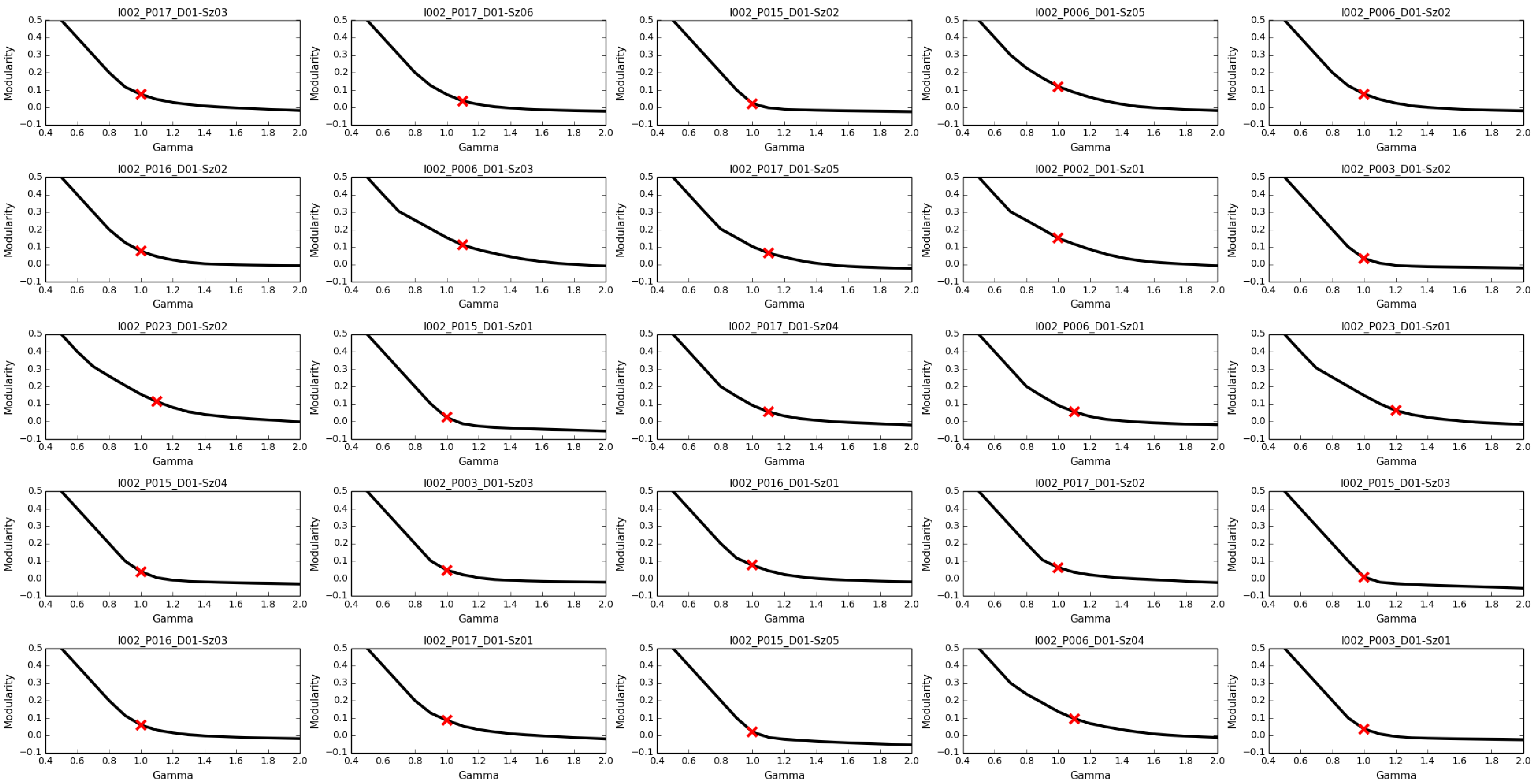}
\caption{\textbf{Structural Resolution Parameter Sweep}. Each graph demonstrates the effect of varying $\gamma$ on the average modularity $\mb{Q}$ over $1000$ optimizations per seizure per patient. The red cross denotes the $\gamma$ chosen for use in the community detection results presented in the main manuscript.\label{figS1}}
\end{figure}

Prior studies have explored a structural resolution parameter limit that represents a trade-off between grouping all nodes into a single community with large modularity and grouping each node into separate communities with low modularity \cite{fortunato2007resolution}. This inverse relationship between modularity and structural resolution has been demonstrated in real-networks \cite{bassett2013robust} suggesting that an even trade-off in modularity and number of communities may occur at the inflection point. To determine an optimal structural resolution parameter for each seizure we computed the inflection point in average modularity $\mb{Q}$ over $1000$ optimizations at each $\gamma$ between $0.5$ and $2.0$ in intervals of $0.1$ (Fig.~S\ref{figS1}).

\section{Network Configuration Communities}
\subsection{Temporal Dispersion of Community Order}
\label{subsec:tempoccur}
The temporal order of configuration communities can be (i) \emph{perfectly cohesive}, where the network reconfigures minimally during an epoch, or (ii) \emph{perfectly random}, where the network reconfigures in every time window. Traditionally, the rate of network reconfiguration has been measured by community flexibility $\mb{F}$, quantified by the number of community assignment changes within an epoch divided by the number of possible changes within the epoch \cite{bassett2011dynamic}. Here we use $\mb{F}$ to measure the temporal dispersion of network configuration patterns, where more ordered community occurrence with low $\mb{F}$ is suggestive of rich temporal structure. Our data show that network configuration communities from dynamic seizure networks demonstrated greater clear cohesive reconfiguration compared to pre-seizure networks that demonstrate more random dispersion of community assignment (Fig.~S\ref{figS2}).

\begin{figure}[b]
\includegraphics[width=\textwidth]{TCommCoh.eps}
\caption{\textbf{Temporal Dispersion of Network Configuration Communities}. Distribution of flexibility $\mb{F}$ over seizure population; $t$-test ($N=25$, $t=11.04$, $p<1\times10^{-14}$).\label{figS2}}
\end{figure}

\subsection{Community Ranking and Distillation}
\label{subsec:communityranking}
\begin{figure}[b]
\includegraphics[width=\textwidth]{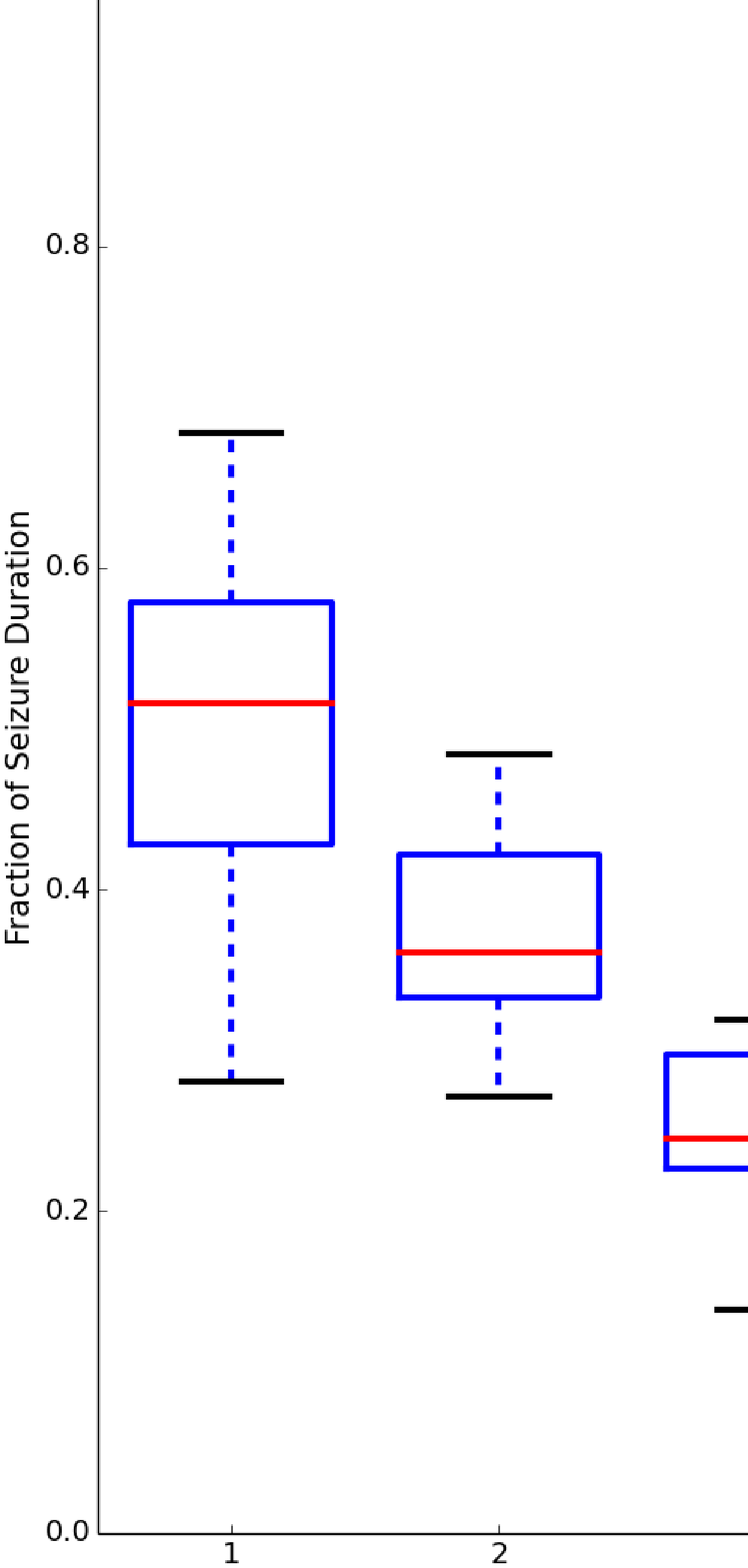}
\includegraphics[width=\textwidth]{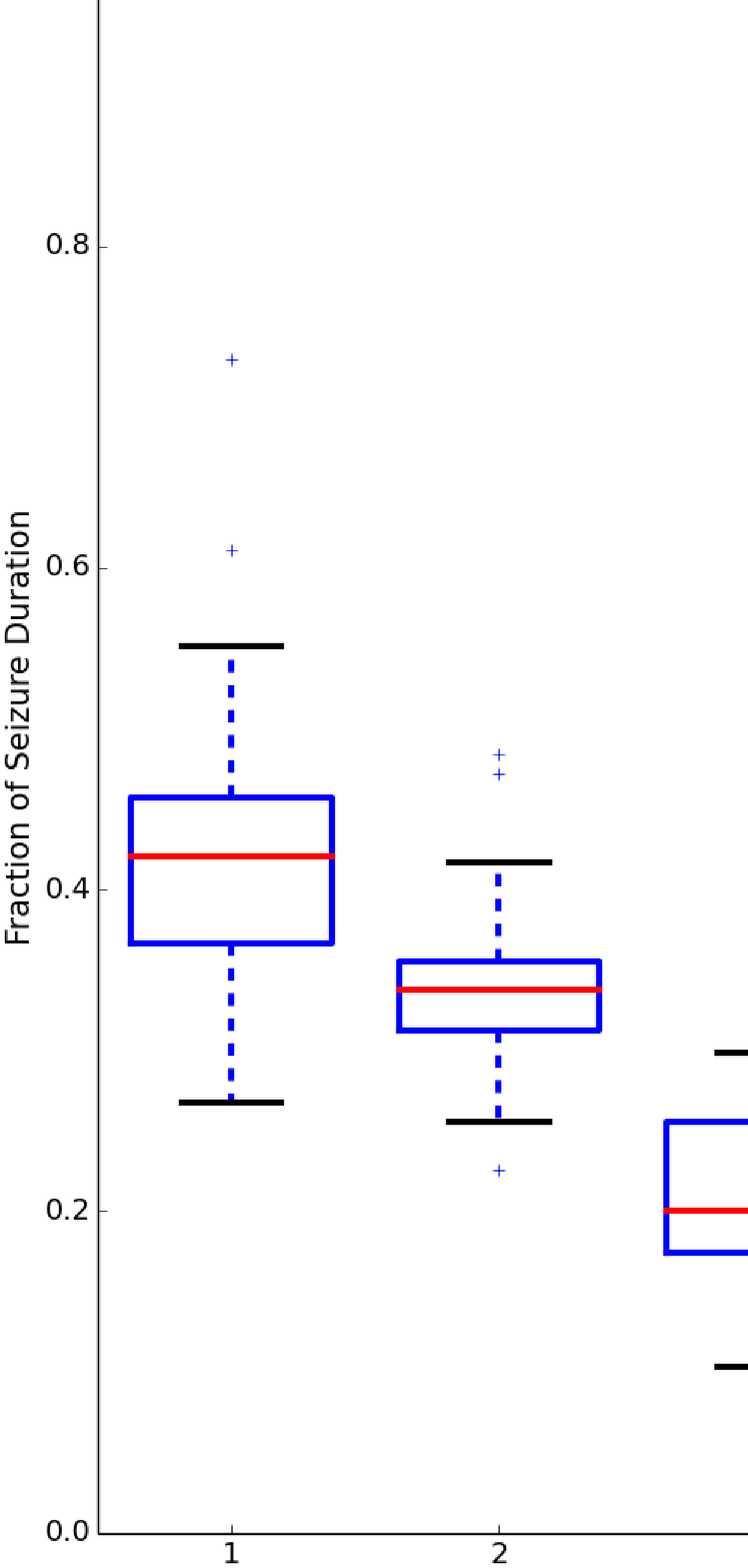}
\caption{\textbf{Size Ranking of Network Configuration Communities}. (\textit{Top}) Pre-seizure network distribution of community size for size-ranked communities over seizure population. (\textit{Bottom}) Seizure network distribution of size-ranked communities over seizure population. \label{figS3}}
\end{figure}

In order to compare configuration communities across seizures, we developed a two-step method to re-assign communities based on temporal occurrence. First, we ranked configuration communities within each pre-seizure network and seizure network by community size; defined by fraction of time windows in an epoch assigned to the community. We then summarized seizure population distributions of community size based on ranked community labels (Fig.~S\ref{figS3}). We found that pre-seizure networks were grouped into at most 3 configuration communities and seizure networks were grouped into at most 6 configuration communities. We also noted that $\approx$90\% or more of the seizure time windows had been assigned to one of the first 3 configuration communities. Therefore, we chose to retain the 3 largest configuration communities from seizure networks for further analysis.

%\bibliography{references.bib}
%\bibliographystyle{pnas.bst}